\documentstyle[12pt]{article}
\newcommand{\zero}{\setcounter{equation}{0}}

\begin{document}

\begin{center}

{\large \bf Gravitational field around a screwed superconducting

cosmic string in scalar-tensor theories}

\vspace{.5 true cm}

V. B. Bezerra$^* \footnote{valdir$@$ufpb.if.br}$ and
C. N. Ferreira$^{\dagger}$\footnote{crisnfer$@$cbpf.br}

$^*$Departamento de F\'{\i}sica, Universidade Federal da Para\'{\i}ba, \\ 
58051-970, J.Pessoa,PB, Brazil

$^{\dagger}$Centro Brasileiro
de Pesquisas F\'{\i}sicas, Rua Dr. Xavier Sigaud 150, Urca
22290-180, Rio de Janeiro,RJ, Brazil

\end{center}



\begin{abstract}
We obtain the solution that corresponds to a screwed superconducting 
cosmic string (SSCS) in the framework of a general scalar-tensor theory 
including torsion.
We investigate the metric of the SSCS in Brans-Dicke theory with torsion
and analyze the case without torsion. We show that in the case with torsion
the space-time background presents other properties different from that in
which torsion is absent. When the spin vanish, this torsion is a 
$\phi$-gradient and then it propagates outside of the string. We investigate 
the effect of torsion on the gravitational force and on the geodesics of a 
test-particle moving around the SSCS. The accretion of matter by wakes 
formation when a SSCS moves with speed $v $ is investigated. 
We compare our results with those obtained for cosmic strings in the framework 
of scalar-tensor theory.
\end{abstract}

\newpage

\section{Introduction }

Topological defects like cosmic strings\cite{Vilenkin,Kibble1,
Vilenkin1,Kibble2} have been studied in different
contexts\cite{Kibble} like, for example, to understand the primordial Universe
and the mecha\-nism of structure formation \cite{Bertschinger,Brandenberger,
SV84,Turok,Vilenkin85} in very early eras.

Cosmic string presents superconducting properties \cite{Witten}
and in this case, it may behaves like both  bosonic (see
Ref.[\cite{Patrick}] and references therein) and fermionic 
strings \cite{Jackiw,Davis99}.
The superconductivity is supposed to be relevant during or very soon
after the phase transition in which the string was formed. 

It has been argued that gravity may be described by a scalar-tensorial 
gravitational field, at least at sufficiently high energy scales. From the 
theoretical point of view, scalar-tensor theories of gravitation, in which 
gravity is mediated by one or several long-range scalar field in addition to 
the usual tensor field present in Einstein's theory, are the most natural 
alternatives to general relativity. In these theories the gravitational 
interaction is mediated by a (spin-2) graviton and by a (spin-0) scalar 
field \cite{Shapiro1,Gaspperini}.

There are compelling suggestions from astrophysical observations that 
Einstein's original description of gravity may require the inclusion of 
hitherto undetected fields of
either gravitational or matter fields at least at the first moments of the 
Universe.
As example, we can mention that the dilaton fields were, certainly, very 
relevant in primordial Universe in comparison with gravitational fields 
but nowadays the scalar contributions is small.

On the other hand,
torsion fields were analyzed in the geometry of a cosmic string whose 
presence could
have been influenced the formation and evolution of  structures in the 
Universe \cite{William}.
Other effect of torsion corresponds to the contribution to neutrino 
oscillations \cite{Adak}.
Due to the role played by the torsion several authors have already 
discussed that torsion
may have been an important element in the early Universe, when the
quantum effects of gravitation were drastically important\cite{Sabbata1,Yishi}.
Also the torsion is important from the phenomenological point of view and it 
may be relevant in cosmology.
This importance is associated with the modifications of kinematic quantities,
like shear, vorticity, acceleration,
expansion and their evolution equations due to the presence of
torsion \cite{Trautman,Stewart,Demianski,Ellis,Palle}.

Therefore, from the previous considerations concerning the importance 
of scalar-tensor theory of gravity and of the theory of gravity which takes
into account the torsion,
we will investigate the superconducting cosmic string in the framework  of
a general scalar-tensor theory with torsion in which the presence of a 
dilaton field and torsion are present and is supposed to persist from the
period of formation of the string on .
We study the formation of the cosmic string wakes in this context and we 
analyze the contribution
of the current carried by the string. Our main
purpose  is to study how the cosmological
effects of long strings are affected by torsion and scalar 
fields in the background
generated by scalar-tensor gravities with torsion as compared with 
general relativity.
We show that the cosmic string wakes moving in space-time with torsion
present effects similar to the wiggly cosmic string, assuming, of course, 
the validity of the Pogosian and Vachaspati conclusion \cite{Pogosian}.

To incorporate the Pogosian and Vachaspati
statement\cite{Pogosian}, we propose that their straight strings
with small-scale structures ($wiggles$) may  resemble the strings
endowed with torsion in our picture (screwed strings). In so
doing, we postulate that the small-scale structures existing in
wiggly strings can be approximately scaled to the geometrical deformation
that torsion produces on ordinary strings. This
premise leads us to the idea that the primordial spectrum of
perturbations in the CMBR, as observed by COBE, may reasonably
be reproduced if one uses the freedom in the parameter-space of
numerical models of structure formation based on wiggly strings.
 
The wakes produced by "wiggly" cosmic strings can result
in an efficient process of formation of large scale
structure and affect the microwave-background
isotropy\cite{Vachaspati91}. In the case of pure scalar-tensor
gravity\cite{Emilia}, the cosmic string wakes present very similar structure
to the wiggly cosmic string in general relativity. In the framework of 
scalar-tensor theory which we are considering, we will show that the 
presence of the torsion amplify this effect.

The shape of the matter (radiation) power spectrum can be obtained
by following the evolution of a network of long ordinary straight
strings interacting with the universe matter (radiation) content.
A string evolves in such a way that its characteristic curvature
radius at time $t$ is $\sim t$ (see Ref.\cite{SV84},
and references therein).
Each string moves with typical speed $v \sim 1$. The translational
motion of a string creates a wedge-shaped {\it wake} behind it with
a deficit angle $8 \pi G \mu$, assuming that $4 \pi G \mu v$ is
greater than the thermal velocity in the network. Under these
conditions the density contrast is $\delta \varepsilon /\varepsilon \sim 1$,
while the wake typical length scale and mass are $\sim t$ and
$M_w \sim 8 \pi G \mu t^3 \delta \varepsilon \sim \mu t$, respectively.
Once the wake forms particles fall into it with transversal
velocity $v_t \sim 4 \pi G \mu$, and acceleration $a_w \sim 2
\pi G  M_w/t^2$. Collisionless particles travel a distance
shorter than the wake width: $v_t^2/2a_w \sim  4 \pi G \mu t$.
In order to be trapped by the wake potential well, baryons must deposit
their energy via a shock. Gravitational instability will force
wakes to grow up to reach masses $\sim 10^{16}$ M$_\odot$ and
length $L_{sc} \sim 20 h^{-1}$ Mpc, characteristic of observed
superclusters of galaxies.

A short description of scalar-tensor theories with  torsion fields
is presented in section 2, where the torsion is considered in a general form.
In section 3 we analyze the specific case where the torsion is a
gradient of the scalar field and  we obtain the superconducting cosmic 
string solution in general scalar-tensor theories with torsion. Sections 4 
and 5 are devoted to some applications and in section 6 we end up with 
some conclusions.

\section{Scalar-tensor theory with torsion\zero}

The scalar-tensor theory with torsion is an extension of
Einstein's general relativity in which a scalar field is coupled
minimally to the gravitational field and a dynamical torsion term
is considered additionally. This coupling
is referred to the Jordan-Fierz frame, where the action takes the
form

\begin{equation}
I=\frac 1{16\pi }\int d^4x\sqrt{{-\tilde g }}\left[ {\tilde \phi}{ \tilde R }-
\frac{\omega(\tilde \phi) }{\tilde  \phi} \partial _\mu \tilde  \phi
\partial^\mu \tilde \phi \right] +I_m ( e_a^{\mu},\Omega_{ab}^
{\hspace{.2 true cm} \mu}, \Psi)\label{acao1},
\end{equation}

\noindent
where $I_{m}( e_a^{\mu},\Omega_{ab}^{\hspace{.2 true cm} \mu}, \Psi)$ is 
the action of the matter which in the general case takes into account all 
fields. The function $ \omega $ in scalar-tensor theory has a 
$\tilde \phi $ dependence  but in Brans-Dicke theory
it is a constant.

The scalar curvature $\tilde R$ is a function of the vierbeins $e_a^{\mu}$
and spin-connections $\Omega_{ab}^{\hspace{.2 true cm} \mu}$\cite{Kim}. In
Einstein-Cartan (EC) theory, $\tilde R$ is given by

\begin{equation}
\tilde R=e^{\mu}_ae^{\nu}_b(\Omega^{ab}_{\hspace{.2 true cm}\mu,\nu} - 
\Omega^{ab}_{\hspace{.2 true
cm}\nu,\mu}+ \Omega^{a}_{\hspace{.2 true cm}c\nu}\Omega^{cb}_
{\hspace{.2 true cm}\mu
}-\Omega^{a}_{\hspace{.2 true cm}c\mu}\Omega^{cb}_{\hspace{.2 true cm}\nu})
\end{equation}

\noindent
where the latin indices $a,b..$ are flat type and the greek
ones $\mu \nu ..$ are world type.

As this point we write the torsion in general form given by

\begin{equation}
S^a_{\hspace{.3 true cm} \mu \nu} =e^a_{\hspace{.2 true cm} \mu, \nu}
-e^a_{\hspace{.2 true cm} \nu, \mu}+ \Omega^a_{\hspace{.2 true cm}
c \nu}e^c_{\mu} - \Omega^a_{\hspace{.2 true cm}
c \mu}e^c_{\nu},
\end{equation}

\noindent
and the field equation for spin connection $\Omega^a_{\hspace{.2 true cm}
c \nu}$ as

\begin{equation}
\tilde \phi (S^{\mu}_{\hspace{.3 true cm} \alpha \beta} +
\delta^{\mu}_{\beta} S^{\lambda}_{\hspace{.3 true cm} \lambda \alpha}
- \delta^{\mu}_{\alpha} S^{\lambda}_{\hspace{.3 true cm} \lambda
\beta}) = 8\pi \sigma^{\mu}_{\hspace{.3 true cm}\alpha \beta} +
\delta^{\mu}_{\alpha}\tilde \phi_{, \beta} - 
\delta^{\mu}_{\beta}\tilde \phi_{,
\alpha},
\end{equation}

\noindent
where $\sigma^{\mu}_{\hspace{.3 true cm}\alpha \beta}$ is the spin
angular momentum tensor defined by

\begin{equation}
\sigma^{\mu}_{\hspace{.3 true cm}\alpha \beta} =
\frac{1}{\sqrt{-g}}(e^a_{\alpha}e^b_{\beta} - e^a_{\beta}
e^b_{\alpha})\frac{\delta}{\delta \Omega^{ab}_{\hspace{.2 true cm}
\mu}}(\sqrt{-g} I_m).
\end{equation}

The contraction of the $\beta$ and  $\mu$ indices leads to

\begin{equation}
\tilde \phi S^{\mu}_{\hspace{.3 true cm} \alpha \beta} = 8
\pi\Sigma^{\mu}_{\hspace{.3 true cm}\alpha \beta} + \frac{1}{2}(
\delta^{\mu}_{\alpha}\tilde \phi_{, \beta} - 
\delta^{\mu}_{\beta}\tilde \phi_{,
\alpha}),
\end{equation}

\noindent
where $\Sigma^{\mu}_{\hspace{.3 true cm}\alpha \beta}$ is given by

\begin{equation}
\Sigma^{\mu}_{\hspace{.3 true cm}\alpha \beta} = 
\sigma^{\mu}_{\hspace{.3 true cm}\alpha \beta}
+ \frac{1}{2}(\delta^{\mu}_{\alpha}\sigma^{\mu}_{\hspace{.3 true cm} \beta
\lambda}- \delta^{\mu}_{\beta} \sigma^{\lambda}_{\hspace{.3 true cm}\alpha
\lambda}).
\end{equation}

It is worth to call attention to the fact that in the scalar 
curvature $\tilde R$, the scalar function $\tilde \phi$ can
act as source of the torsion field. Thus, in the absence of spins, the
torsion field may be generated by a non-spin term, the gradient of
the scalar field and therefore, in the absence of spin the torsion does not
vanish. Therefore, it can propagates with the scalar field and thus we can 
write the torsion as

\begin{equation}
S_{\mu \nu}^{\hspace{.3 true cm} \lambda} =
\left(\delta^{\lambda}_{\mu}
\partial_{\nu}\tilde  \phi -\delta^{\lambda}_{\nu} \partial_{\mu}
\tilde \phi\right)/2\tilde \phi \label{torc1}.
\end{equation}

The most general affine connection $\Gamma_{\lambda \nu}^{\hspace{.3 true
cm} \alpha}$  written  in terms of the contortion tensor
$K_{\lambda \nu}^{\hspace{.3 true cm}\alpha}$ is

\begin{equation}
\Gamma_{\lambda \nu}^{\hspace{.3 true cm} \alpha} =
\{^{\alpha}_{\lambda \nu}\} +
 K_{\lambda \nu}^{\hspace{.3 true cm}\alpha}, \label{kont1}
\end{equation}

\noindent
where the quantity  $ \{^{\alpha}_{\lambda \nu}\}$ is the Christoffel  symbol
computed from the metric tensor $g_{\mu \nu}$ and the contortion tensor
$K_{\lambda \nu}^{\hspace{.3 true cm}\alpha}$ can be written
in  terms of the torsion field as:

\begin{equation}
K_{\lambda \nu}^{\hspace{.3 true cm}\alpha} =
-\frac{1}{2}(S_{\lambda \hspace{.2 true cm} \nu}^{\hspace{.1 true
cm} \alpha} + S_{\nu \hspace{.2 true cm} \lambda}^{\hspace{.1 true
cm} \alpha} - S_{\lambda \nu}^{\hspace{.3 true cm} \alpha})
\end{equation}

In this case the scalar curvature $\tilde R$ given by (\ref{acao1}) in 
the Jordan-Fierz frame can be written as

\begin{equation}
\tilde R = \tilde R(\{\}) + \epsilon \frac{\partial_{\mu} 
\tilde \phi \partial^{\mu} \tilde \phi}{\tilde \phi^2},
\end{equation}

\noindent
where $\tilde R(\{\})$ is the Riemann scalar curvature in the Jordan-Fierz 
frame and $\epsilon $ is the torsion coupling constant \cite{Gaspperini}.

Although action (\ref{acao1}) shows explicitly this scalar-tensor gravity's
character, we will adopt, for technical reasons, to work in the
conformal (Einstein) frame in which the kinematic terms of the scalar and 
the tensor fields do not mix and the action is given by

\begin{equation}
{I} = \frac{1}{16\pi G} \int d^4x \sqrt{-g} \left[ R -
2g^{\mu\nu}\partial_{\mu}\phi\partial_{\nu}\phi \right]
+ I_{m}[\Psi_m,\Lambda^2(\phi)g_{\mu\nu}] ,\label{acao2}
\end{equation}

\noindent
where $g_{\mu\nu}= e^a_{\mu}e^b_{\nu} \eta_{ab}$ is a pure rank-2 tensor 
in the Einstein frame and $R$ is the curvature scalar such that

\begin{equation}
R = R(\{\}) + 4\epsilon \alpha(\phi)^2 \partial_{\mu} 
\phi \partial^{\mu} \phi \label{R}
\end{equation}

It is interesting to note that action (\ref{acao2}) is
obtained from (\ref{acao1}) by a conformal transformation

\begin{equation}
\tilde{g}_{\mu\nu} = \Lambda^2(\phi)g_{\mu\nu} , \label{conform}
\end{equation}

\noindent
and by a redefinition of the quantity

\[G\Lambda^2(\phi) = \tilde{\phi}^{-1}\]

\noindent
which shows up that any gravitational
phenomena will be affected by the variation of the gravitation ``constant"
$G$ in the scalar-tensorial gravity, and finally, by introducing a new 
parameter

\[\alpha^2 \equiv \left( \frac{\partial \ln \Lambda(\phi)}{\partial
\phi} \right)^2 = [2\omega(\tilde{\phi}) + 3]^{-1} ,\]

\noindent
which can be interpreted as the (field-dependent) coupling
strength between matter and the scalar field.

In order to make our calculations as general as possible, we
choose not to specify the factors $\Lambda(\phi)$ and
$\alpha (\phi)$
(the field-dependent coupling strength between matter and
the scalar field), leaving them as arbitrary functions of the scalar field.

In the conformal frame, the Einstein equations are modified.  A
straightforward calculation shows that they turn into

\begin{eqnarray}
R_{\mu \nu }&=&2\kappa \partial _\mu \phi \partial
_\nu \phi +8\pi G(T_{\mu \nu} - \frac{1}{2}g_{\mu
\nu}T).\label{escalar1}\\
G_{\mu\nu} & = & 2\kappa \partial_{\mu}\phi\partial_{\nu}\phi -
\kappa g_{\mu\nu}g^{\alpha\beta}\partial_{\alpha}\phi\partial_{\beta}
\phi + 8\pi G T_{\mu\nu}\label{eins},
\end{eqnarray}

\noindent
where $\kappa $ is  $\phi$-function defined by

\begin{equation}
\kappa(\phi) = 1- 2 \epsilon \alpha(\phi)^2
\end{equation}

\noindent
which has two contributions: one  given by scalar-tensor term and 
the other by the torsion.
We note that the last equation brings a new information and shows
that the matter distribution behaves as a source for $\phi$
and $g_{\mu\nu}$ as well. The energy-momentum tensor is defined
as usual

\begin{equation}
T_{\mu\nu} \equiv \frac{2}{\sqrt{-g}}\frac{\delta I_m}
{\delta g_{\mu\nu}} ,\label{tensor}
\end{equation}

\noindent
but in the conformal frame it is no longer conserved
$\nabla_{\mu}T^{\mu}_{\nu} = \alpha(\phi)T\nabla_{\nu}\phi$. It
is clear from transformation (\ref{conform}) that we can relate quantities
from both frames in such a way that $\tilde{T}^{\mu\nu} =
\Lambda^{-6}(\phi)T^{\mu\nu}$ and $\tilde{T}^{\mu}_{\nu} =
\Lambda^{-4}(\phi)T^{\mu}_{\nu}$.

In the scalar-tensor theory, the Einstein equations are modified by
the presence of the field $\phi$ and are obtained by applying the
variational  principle to (\ref{acao2}) with $R$ given by (\ref{R}).
The equation for $\phi $
reads as follows:

\begin{equation}
\Box _g\phi =-4\pi G\alpha (\phi )T  \label{eq1},
\end{equation}

\noindent
where
\begin{equation}
\Box_g \phi = \frac{1}{\sqrt{-g}}\partial_{\mu}
\left[\sqrt{-g}\kappa \partial^{\mu} \phi\right].
\end{equation}

It brings some  new information because it doesn't
appear in general relativity, and shows us that a matter
distribution in the space behaves like a source for $\phi $, and,
as usual, for $g_{\mu \nu}$ as well.
Up to now, we have dealt with the purely gravitational sector, but in
what follows, we will introduce the action for the matter that
describes a cosmic string.

\section{Superconducting cosmic string in scalar-tensor theory   \zero}

In order to describe the simplest superconducting cosmic string in a 
scalar-tensor theory, we require   the matter action to carry a pair of 
complex scalar and gauge
fields, in an  Abelian Higgs  model  whose action is  given by

\begin{equation}
I_m = \int d^4x \sqrt{{\tilde g}}[-\frac{1}{2}D_{\mu}\Phi
(D^{\mu}\Phi)^* - \frac{1}{2}D_{\mu} \Sigma (D^{\mu}
\Sigma)^* - \frac{1}{4}F_{\mu \nu}F^{\mu \nu} - \frac{1}{4}H_{\mu
\nu}H^{\mu \nu} - V(|\Phi|, |\Sigma |)],\label{acao3}
\end{equation}

\noindent
where $ D_{\mu} \Sigma = (\partial_{\mu} + ie A_{\mu}) \Sigma $ and
$D_{\mu} \Phi = ( \partial_{\mu} + i qC_{\mu})\Phi $ are the covariant
derivatives. The reason why the gauge fields do not minimally couple to
torsion is well discussed in Refs.\cite{Gaspperini,Hehl}.  The field
strengths are defined as usually as $ F_{\mu \nu} =
\partial_{\mu}A_{\nu} - \partial_{\nu} A_{\mu}$ and $ H_{\mu \nu} =
\partial_{\mu}C_{\nu} - \partial_{\nu} C_{\mu} $, with $A_\mu$ and
$C_\nu$ being the gauge fields and $V(|\Phi |, |\Sigma)|$.

This action given by Eq.(\ref{acao3}) has a $U(1)' \times U(1)$ symmetry, 
where the $U(1)' $ group, associated with the $\varphi$-field, is broken by 
the vacuum and gives rise to vortices of the Nielsen-Olesen type\cite{Nielsen}

\begin{equation} \begin{array}{ll} \Phi = \varphi(r )e^{i\theta},\\
C_{\mu} = \frac{1}{q}[P(r) - 1]\delta^{\theta}_{\mu},
\end{array}\label{vortex1} \end{equation}

\noindent
in which $(t,r,\theta,z)$ are usual cylindrical coordinates.
The boundary conditions for the fields $\varphi(r) $ and $P(r)$ are 
the same as those of
ordinary cosmic strings\cite{Nielsen}, namely

\begin{equation}
\begin{array}{ll}
\begin{array}{ll}
\varphi(r) = \eta & r \rightarrow
\infty \\
\varphi(r) =0 & r = 0 \end{array}&
\begin{array}{ll}
P(r) =0 & r \rightarrow \infty \\
P(r) =1 & r= 0.  \end{array}
\end{array} \label{config1}
\end{equation}

The other $U(1)$- symmetry, that we associate with electromagnetism, acts
on the  $\Sigma $-field. This symmetry is not broken by the vacuum;
however, it is broken in the interior of the defect. The $\Sigma
$-field in the string core, where it acquires an expectation value, is
responsible for a bosonic current being carried by the gauge field
$A_{\mu}$. The only non-vanishing components of the gauge fields are
$A_z(r)$ and $A_t(r)$, and the current-carrier phase may be expressed as
$\zeta(z,t) = \omega_1 t - \omega_2z$. Notwithstanding, we focus only
on the magnetic case \cite{Patrick}. Their configurations are defined
as:

\begin{equation}
\begin{array}{ll}
\Sigma = \sigma(r)e^{i\zeta(z,t)},\\
A_{\mu} = \frac{1}{e}[A(r) - \frac{\partial \zeta(z,t)}{\partial
z}]\delta_{\mu}^{z},
\end{array}
\label{vortex2}
\end{equation}

\noindent
because of the rotational symmetry of the string itself. The fields
responsible for the cosmic string superconductivity have
the following boundary conditions

\begin{equation}
\begin{array}{ll}
\begin{array}{ll}
\frac{d}{d r}\sigma(r) = 0 &r=0 \\
\sigma(r) = 0 & r \rightarrow \infty \end{array} &
\begin{array}{ll}
A(r) \neq 0 & r \rightarrow
\infty \\
A(r) = 1 & r = 0.
\end{array}
\end{array}
\label{config2}
\end{equation}

The potential $V(\varphi, \sigma)$ triggering the spontaneous
symmetry breaking can be fixed by:

\begin{equation} V(\varphi, \sigma) = \frac{\lambda_{\varphi}}{4} (
\varphi ^2 - \eta^2)^2 + f_{\varphi \sigma}\varphi ^2\sigma ^2 +
\frac{\lambda_{\sigma}}{4}\sigma ^4 -
\frac{m_{\sigma}^2}{2}\sigma^2 , \end{equation}

\vspace{.5 true cm}

\noindent
where $\lambda_{\varphi}$, $\lambda_{\sigma}$, $f_{\varphi \sigma}$ and 
$m_{\sigma}$
are coupling constants. Constructed in this way, this potential 
possesses all the ingredients that make it viable to generate the 
formation of a superconducting cosmic string,
as it is well-stablished. In addition, it is extended to include a new term
describing the interaction with the torsion field. The presence of this
interaction term does not affect the occurrence of the string ground states.
However, it adds with a torsion density in the string core due to the coupling
with the charged particle flux.

Let us consider a cosmic string in a cylindrical coordinate
system, $(t,r, \theta ,z)$, ($r \geq 0$ and $0 \leq \theta < 2 \pi
$) with the metric defined in Einstein frame as 

\begin{equation}
ds^2 = e^{2(\gamma - \psi)}(-dt^2 + dr^2 ) + \beta^2
e^{-2\psi}d\theta^2 + e^{2\psi}dz^2 \label{metric1},
\end{equation}

\noindent where $\gamma, \psi$ and  $\beta$ depend only on $r$. We
can find the relations between the parameters of the metric
through
Einstein equations as given by Eq (\ref{eins}). Then, in the space-time
with the metric defined in Eq (\ref{metric1}), these equations are

$$
\beta'' = 8 \pi G \beta ( T^t_t + T^r_r) e^{2(\gamma -
\psi)}
$$ 

\begin{equation}
(\beta \gamma')' = 8 \pi G \beta ( T^r_r +
T^{\theta}_{\theta}) e^{2(\gamma - \psi)},\label{eqq1}
\end{equation}

$$
(\beta \psi')'= 4 \pi G \beta ( T^t_t +
T^r_r + T^{\theta}_{\theta} -T^z_z)e^{2(\gamma -
\psi)},
$$

\noindent
and the $\phi$ - equation given by (\ref{eq1}) is

\begin{equation}
(\beta \kappa
\phi')'= 4\pi G \beta T \alpha(\phi) e^{2(\gamma -
\psi)},\label{efi1}
\end{equation}
where $(')$ denotes ``derivative with respect to r".

The non-vanishing components of the energy-momentum tensor are

\vspace{.5 true cm}

\begin{eqnarray}
T^{t}_{t} & = &  - \frac{1}{2}\Lambda^{2}(\phi) \{
e^{2(\Psi - \gamma)}(\varphi'^{2} + \sigma'^{2}) + \frac{e^{2\Psi}}
{\beta^2}\varphi^2P^2 + e^{-2\Psi}\sigma^2A^2  \nonumber \\
& & + \Lambda^{-2}(\phi)e^{-2\gamma}(\frac{A'^2}{4\pi e^2}) +
\Lambda^{-2}(\phi)\frac{e^{2(2\Psi - \gamma)}}{\beta^2}(\frac{P'^2}
{4\pi q^2}) + 2\Lambda^2(\phi)V(\varphi,\sigma) \} \nonumber \\
T^{r}_{r} & = &  \frac{1}{2}\Lambda^2(\phi) \{ e^{2(\Psi - \gamma)}
(\varphi'^2 + \sigma'^2) - \frac{e^{2\Psi}}{\beta^2} \varphi^2P^2 - 
e^{-2\Psi}\sigma^2A^2 \nonumber \\
& & + \Lambda^{-2}(\phi)e^{-2\gamma}(\frac{A'^2}{4\pi e^2}) + \Lambda^{-2}(\phi)
\frac{e^{2(2\Psi - \gamma)}}{\beta^2}(\frac{P'^2}{4\pi q^2}) - 2\Lambda^2(\phi)
V(\varphi,\sigma) \} \nonumber \\
T^{\theta}_{\theta} & =  & - \frac{1}{2}\Lambda^2(\phi) \{ e^{2(\Psi - 
\gamma)}(\varphi'^2 + \sigma'^2) - \frac{e^{2\Psi}}{\beta^2} \varphi^2P^2 +
e^{-2\Psi}\sigma^2A^2  \\
& & + \Lambda^{-2}(\phi)e^{-2\gamma}(\frac{A'^2}{4\pi e^2}) - \Lambda^{-2}(\phi)
\frac{e^{2(2\Psi - \gamma)}}{\beta^2}(\frac{P'^2}{4\pi q^2}) + 2\Lambda^2(\phi)
V(\varphi,\sigma) \} \nonumber \\
T^{z}_{z} & = & - \frac{1}{2}\Lambda^2(\phi) \{ e^{2(\Psi - \gamma)}
(\varphi'^2 + \sigma^2) + \frac{e^2\Psi}{\beta^2}\varphi^2P^2 - e^{-2\Psi}\sigma^2
A^2 \nonumber \\
& & - \Lambda^{-2}(\Phi)e^{-2\gamma}(\frac{A'^2}{4\pi e^2}) + \Lambda^{-2}(\phi)
\frac{e^{2(2\Psi -\gamma)}}{\beta^2}(\frac{P'^2}{4\pi q^2}) +  2\Lambda^2(\phi)
V(\varphi,\sigma) \nonumber \}
\end{eqnarray}

\vspace{1 true cm}

The external solutions of the Eq.(\ref{eqq1}) have the same form
of the scalar tensor  theory \cite{Cris},
 but the $\phi$-solution is different and comes from

\begin{equation}
\phi'= \kappa^{-1} \frac{\lambda }{r}.
\end{equation}

\noindent
This implies that

\begin{equation}
R=2(\psi''+\frac{1}{\beta}\psi' - \phi'^2 +
\frac{m}{\beta^2})e^{2(\psi-\gamma)}=  2\phi'^2e^{2(\psi-\gamma)}
\end{equation}

\noindent
is different from the result obtained in the framework of pure scalar-tensor
theories of gravity\cite{Cris}.

Let us make an estimation of the order of magnitude of the
correction  induced by
$ \kappa^{-1}\lambda $. It is very illustrative to consider a
particular form for the arbitrary function $\Lambda(\phi)$,
corresponding to the Brans-Dick theory, $\Lambda=e^{\alpha
\phi}$, with $\alpha^2 = \frac{1}{2w +3}$, ($w$=cte). Thus, for this case,

\begin{equation}
\phi'= \lambda \frac{ (w +\frac{3}{2})}{(\tilde w + \frac{3}{2}) r}
\end{equation}

\noindent
where $\tilde \omega = \omega - \epsilon $.
Using the values for $w$ such that $w > 2500$ (consistent with
solar system experiments made by Very Baseline Interferometry
(VLBI) \cite{Eubanks}, we have that  $\omega >>> \epsilon$ and 
thus $\frac{\lambda (w +\frac{3}{2})}{(\tilde w + \frac{3}{2})} \sim
\lambda $. Therefore, the external solution in the Brans-Dicke theory in this 
limit is the
same as in the case of the superconducting cosmic
string in scalar-tensor theory \cite{Cris}.

\vspace{.5 true cm}

The external metric for the SSCS takes, thus, the form

\begin{equation}
ds^2 =  \left( \frac{r}{r_0} \right)^{-2n} W^2(r) \left[
\left( \frac{r}{r_0}\right)^{2m^2} (-dt^2 +dr^2) + 
B^2r^2d\theta^2 \right] + \left( \frac{r}{r_0} \right)^{2n}
\frac{1}{W^2(r)} dz^2 \label{m8},
\end{equation}

\noindent
with $W(r) = [(r/r_0)^{2n} + p]/[1+p]$ and the parameters $n, \lambda $ and
$m$ are given by $n^2 = \kappa^{-1}\lambda^2 + m^2 $, with $\kappa^{-1}$ 
constant in the Brans-Dicke theory.

\vspace{.5 true cm}

Now, let us find the solutions for a SSCS by considering the weak field 
aproximation ( for a review of the procedure see Ref.\cite{Cris}). 
To do this let us assume that the metric $g_{\mu\nu}$ and the scalar
field $\phi$ can be written as:

\begin{equation}
\begin{array}{ll}
g_{\mu\nu} = \eta_{\mu\nu} + h_{\mu\nu} ,\\
\Lambda(\phi) = \Lambda(\phi_0) +  \Lambda'(\phi_0)\phi_{(1)}, \\
T_{\mu \nu} = T_{(0)\mu\nu} + T_{(1)\mu \nu},\\
\phi = \phi_0 + \phi_{(1)},
\end{array}
\end{equation}

\noindent
where $\Lambda'(\phi_0) =\Lambda(\phi_0) \alpha(\phi_0)$, 
$\eta_{\mu\nu} = diag(-,+,+,+)$ is the Minskowski metric
tensor and $\phi_0$ is a constant.

\noindent

The energy per unit length $U$, the tension per unit length $\tau $ and
the current density $I$, are given, respectively, by

\begin{equation}
U  = -2 \pi \int_0^{r_0} T^t_t r dr ;
\end{equation}

\begin{equation}
\tau = - 2\pi \int_0^{r_0} T^z_z r dr ;
\end{equation}

\noindent
and

\begin{equation}
I = 2\pi e \int_0^{r_0} r dr \sigma^2 A
\end{equation}

In the case of a space-time with torsion, we can find the matching 
conditions using the fact
 that $[\{^\alpha_{\mu\nu}\}]_{_{r=r_0}}^{(+)}=
[\{^\alpha_{\mu\nu}\}]_{_{r=r_0}}^{(-)}$, and the metricity constraint
$[\nabla_{\rho}g_{\mu \nu}]_{_{r=r_0}}^{+}= [\nabla_{\rho}g_{\mu
\nu}]_{_{r=r_0}}^{-}= 0$. Thus, we find the following continuity conditions

\begin{eqnarray}
&[g_{\mu \nu}]_{_{r=r_0}}^{(-)} =
[g_{\mu \nu}]_{_{r=r_0}}^{(+)}, \nonumber \\
&[\frac{\partial g_{\mu
\nu}}{\partial x^{\alpha }}]_{_{r=r_0}}^{(+)} + 2 [g_{\alpha \rho}
K_{(\mu \nu)}^{\hspace{.3 true cm} \rho}]_{_{r=r_0}}^{(+)} =
[\frac{\partial g_{\mu \nu}}{\partial x^{\alpha}}]_{_{r=r_0}}^{(-)} +
2 [g_{\alpha \rho} K_{(\mu \nu)}^{\hspace{.3 true cm}
\rho}]_{_{r=r_0}}^{(-)}, \label{junc1}
\end{eqnarray}

\noindent
where $(-)$ represents the internal region and $(+)$ corresponds to the
external region around $r = r_0$. In analyzing the junction conditions
we notice that the contortion contributions  appear neither in
the internal nor in the external regions, differently from the results
obtained in Refs.( \cite{Kopczynski,Volterra}).

In order to compare the external solutions we demand a linearisation of these
ones since they are exact solutions. Then, we make a change of
variable $r \rightarrow \rho$, \cite{Cris1}, such that

\begin{equation}
\rho = r \left[ 1 + \tilde{G}_{0} (4U + I^2) - 
4\tilde{G}_{0} U \ln \frac{r}{r_0} - 
2\tilde{G}_{0}I^2 \ln^2 \frac{r}{r_0}\right] ,
\end{equation}

\noindent
and, in this way, we have that at first order in $\tilde G_0$,
the $\phi_{(1)}$-solution is given by

\begin{equation}
\phi_{(1)} = 2\tilde{G}_{0} \kappa^{-1} \alpha(\phi_0) (U+
\tau -I^2)\ln \frac{\rho}{r_0} .
\end{equation}

\noindent
where we used Eq.(\ref{eq1}) and the fact that $\tilde G_0 \equiv G  
\Lambda^2(\phi_0)$.

Doing the  identification of the coefficients of both linearised metrics, 
we finally obtain

\begin{eqnarray}
m^2 & = & 4\tilde{G}_0 I^2 \nonumber \\
B^2 & = & 1 - 8\tilde{G}_0 (U+\frac{I^2}{2}) \nonumber \\
\lambda & = & 2\tilde{G}_0 \alpha(\phi_0) (U+\tau-I^2) \nonumber \\
p & = & 1 + \tilde{G}^{1/2}_0 (U - \tau -I^2) .
\end{eqnarray}

Using the same procedure to obtain the superconducting cosmic string 
solution in the framework of the scalar-tensor theories \cite{Cris},
we find the metric in Einstein frame as

\begin{eqnarray}
ds^2 & = & \left\{ 1 + 4\tilde{G}_0 \left[ I^2\ln^2 \frac{\rho}{r_0} + 
(U-\tau +I^2)\ln
\frac{\rho}{r_0}\right] \right\} (-dt^2 + d\rho^2) \nonumber \\
& & + \left\{ 1 - 4\tilde{G}_0 \left[ I^2 \ln^2\frac{\rho}{r_0} +
(U-\tau -I^2) \ln \frac{\rho}{r_0} \right]\right\} dz^2 \label{m5} \\
& & + \rho^2 \left[ 1 - 8\tilde{G}_0 (U+ \frac{I^2}{2}) + 
4\tilde{G}_0 (U-\tau -I^2)\ln \frac{\rho}{r_0} + 
4\tilde{G}_0 I^2 \ln^2\frac{\rho}{r_0} \right] d\theta^2 \nonumber .
\end{eqnarray}

Note that this metric corresponds to the same one obtained in the case of pure
scalar-tensor theories. The reason for this coincidence is that the 
contribution due to the torsion comes out only in the third order and 
therefore it does not appear in the linearized solution we have considered. 
Otherwise, some new physical effects appears as we shall see in the next 
section.

The deficit angle associated with the space-time given by 
metric (\ref{m5}) is

\[\Delta\theta = 2\pi \left[ 1 - 
\frac{1}{\sqrt{g_{\rho\rho}}}\frac{d}{d\rho} 
\sqrt{g_{\theta\theta}} \right] ,
\]

\noindent
which can be written as

\begin{equation}
\Delta\theta = 4\pi \tilde{G}_0 (U +\tau + I^2) .
\end{equation}

\section{Particle deflection near a SSCS \zero}

In this section we study the geodesic equation in the space-time 
under consideration. To do this,
we have to work with the metric given by Eq.(\ref{m5}) in the Jordan-Fierz
frame. Then, if we consider a section perpendicular
to the string, i. e., $dz=0$, then we have

\begin{equation} ds^2_{\perp}= \Lambda(\phi_0)^2(1-h_{tt})[-dt^2 + 
dr^2 + (1- b) r^2d\theta^2], \end{equation} \label{lin1}

\noindent
where $h_{tt}$ is

\begin{equation}
h_{tt} = -4\tilde G_0\{I^2 (\ln(\frac{\rho}{r_0}))^2 + 
[\alpha_+U - \alpha_-(\tau - I^2) ]\ln(\frac{\rho}{ r_0}\}\label{htt},
\end{equation}

\noindent
with $b$ and $\alpha $ being given by

\begin{equation}
b = 8\tilde{G}_0 \left[U+ \frac{I^2(1+ 2\ln \frac{\rho}{r_0} ) }{2} \right],
\end{equation}

\noindent
and

\begin{equation}
\alpha_{\pm}= [1 \pm \frac{1}{2}\kappa^{-1}\alpha^2(\phi_0)].
\end{equation}

We know that when the string possesses current there appear
gravitational forces.  We shall consider the effect that torsion
plays on the gravitational force generated by a SSCS on a particle
moving around the defect, assuming that the particle has no
charge. Let us consider the particle with speed $|{\bf v}| \leq 1 $,
in which case the geodesic equation becomes

\begin{equation}
\frac{d^2x^i}{d\tau^2} + \Gamma_{tt}^{\hspace{.2 true cm} i} =0, \label{geo1}
\end{equation}

\noindent
where  $i$ is the spatial coordinate index and the connection can be written as
in Eq.(\ref{kont1}), with  the non-vanish terms given by

\begin{equation}
\Gamma^{\hspace{.2 true cm}i}_{(tt)} = 
\left\{ ^{\hspace{.1 true cm} i}_{tt}\right\} +
K^{\hspace{.2 true cm}i}_{(tt)}
\end{equation}

\noindent
with $K^{\hspace{.2 true cm}i}_{(tt)}$ being the  contortion
which is symmetric in the two-first indices. The only  non-vanish part is

\begin{equation}
K^{\hspace{.2 true cm}r}_{(tt)} = \frac{\tilde \phi'}{2\tilde
\phi}\sim -\alpha(\phi_0) \phi_{(1)}'=
-2\frac{1}{\rho}\tilde{G}_{0}\kappa^{-1} \alpha^2(\phi_0) (U+\tau -I^2).
\end{equation}

The gravitational acceleration  around the string gets the form

\begin{equation}
a =  \nabla h_{tt} - 2\frac{\tilde{G}_{0} \kappa^{-1}\alpha^2(\phi_0) 
(U+\tau -I^2)}{\rho},
\end{equation}

\noindent
with $g_{tt}= -1+ h_{tt} $ in Eq.(\ref{m5}). Therefore, the torsion 
contribution to the force is

\begin{equation}
f_{_{tors}}= -\frac{2 m}{\rho}\tilde{G}_{0}\kappa^{-1} \alpha^2(\phi_0) 
(U+\tau -I^2)
\end{equation}

We also note that the gravitational pull is related to the $h_{tt}$ component
that has  explicit dependence on the torsion, as shown in Eq.(\ref{htt}).
>From the last equation, the  force that the SSCS exerts on a test particle
can be explicitly written as

\begin{equation}
f = -\frac{m}{\rho} \left[ 4\tilde G_{0}I^2\left(1 + \frac{(U-
\tau )}{I^2} + 2\ln(\rho/r_0) \right) + 
4\tilde{G}_{0} \kappa^{-1}\alpha^2(\phi_0) (U+\tau -I^2)\right]. \label{force}
\end{equation}

A quick glance at the last equation allows us to understand the
essential role played by torsion  in the context of the present
formalism. If torsion is present, even in the case in which the
string has no current, an attractive gravitational force appears.
In the context of the SSCS, torsion acts such to enhancer the
force that a test particle feels outside the string. This peculiar
fact may have meaningful astrophysical and cosmological effects,
as for example, influencing the process of formation of structures.

\section{Large-scale structure formation \zero}

Let us consider the deflection of particles moving past the string.
Assuming for simplicity that the direction of  propagation is
perpendicular to the string, we can write the metric (\ref{lin1}) in
terms of Minkowskian coordinates  in
the form

\begin{equation}
ds^2 = (1 - h_{00})[dt^2 -dx^2 -dy^2]\label{lin2}
\end{equation}

In last section we concluded that in a space-time with torsion
there is a change in the geodesics due to the presence of the symmetric 
part of the contortion (\ref{geo1}). To study the formation of a wake
behind a moving screwed cosmic string, we will first consider the rest 
frame of the string with a velocity $v$ in the x direction. In this 
situation, all components of geodesic equations are

\begin{equation}
\frac{d^2x^i}{d\tau^2} + 
\Gamma_{(\mu \nu)}^{\hspace{.2 true cm} i}u^{\mu}u^{\nu} =0, \label{geo2}
\end{equation}

\noindent
whose linearized forms are given by

\begin{equation}
2\ddot{x} = -(1-\dot{x}^2-\dot{y}^2)\partial_xh_{tt} + 
(1-\dot{y}^2)\alpha(\phi_0)\partial_x\phi_{(1)}\label{y0},
\end{equation}

\begin{equation}
2\ddot{y} = -(1-\dot{x}^2-\dot{y}^2)\partial_yh_{tt} + 
(1-\dot{x}^2)\alpha(\phi_0)\partial_y\phi_{(1)} \label{y},
\end{equation}

\noindent
where $h_{00}$ is given by (\ref{htt}) and the overdot
denotes derivative with respect to $t$.
We can analyze Eqs.(\ref{y0}) and (\ref{y}) in which concerns the
contribution coming from the last term.

We need only consider terms of first order in $\tilde G_0$,
in  which case (\ref{y}) can integrated over the
unperturbed trajectory $x =vt$, $y=y_0$. Then,
we can transform to the frame in which the string
has a velocity $v$. Then, particles enter the
wake with a transverse velocity

\begin{equation}
v_t = 4\pi \tilde G_0(U +\tau + I^2)v \gamma +
\frac{4\pi\tilde G_0 \kappa^{-1} \alpha^2(\phi_0)(U + 
\tau - I^2)}{v \gamma}\label{transversal}
\end{equation}

The first term is the usual velocity impulse of the particles due to
the deficit angle. The second term contains the contribution due to the 
torsion.

The motion of the string creates the wakes, then we
can compute the characteristic total mass of the
large structures formed in this way as

\begin{equation}
M_w=2 v_t \varepsilon t^3
\end{equation}

The gravitational acceleration in the field of the wake is

\begin{equation}
a_w = 2 \pi G M_{w}/t^2.
\end{equation}

Thus the wake width, i.e., the scale of the largest
structures in the universe, turns out to be

\begin{equation}
\Delta l \equiv \frac{v^2_t}{2 a_w} \sim v_t t  .
\end{equation}

From these equations, and using $t_{eq} = 4\times 10^{10}
(\Omega h)^{-2}s$  for the universe age at the radiation-matter
equilibrium phase, we can determine the main characteristics of
the gravitational perturbations induced by wakes formed from
SSCS.

\section{Conclusion}

It is possible that torsion had a physically relevant role during the early 
stages of the Universe's evolution. Along these lines, torsion fields may 
be potentially sources of dynamical stresses which, when coupled to other 
fundamental fields (i.e., the gravitational and scalar fields), might have 
performed an important action during the phase transitions leading to 
formation of topological defects such as the SSCS here we have considered. 
Therefore, it seems an important issue to
investigate basic models and scenarios involving cosmic defects within the 
context of scalar-tensor theories with torsion. We showed that torsion as 
well as scalar fields has a small but non-negligible contribution to 
the geodesic equation obtained from the contortion term and from the scalar 
fields, respectively. From a physical point of view,
these contributions, certainly, are important, and must be considered.
The motivation to consider this scenario comes form the fact that scalar-tensor
gravitational fields are important for a consistent description of gravity, at
least at sufficiently high energy scales, and that torsion as well can induce
some physical effects and could be important at some energy scale, as for 
example, in the low-energy limit of string theory.

As we showed in this work, massless particles (such as photons) will be 
deflected by an angle $\Delta \theta = 4 \pi \tilde G_{(0)}(U+
\tau +I^2)$. From the observational point of view, it would be
impossible to distinguish a screwed string from its general
relativity partner, just by considering effects based on deflection
of light (i.e., double image effect, for instance). On the other
hand, trajectories of massive particles will be affected by the
torsion coupling (which is generated by a space-time with torsion) 
\cite{Cris1,Kleinert2000}. 
 
If the string is moving with normal velocity, $v$, through matter,
a transversal velocity appears which is given by Eq.(\ref{transversal}). 
It is worth to call attention to the fact that there exists, in this case,
a new contribution to the transversal velocity given by 
$v_t = \frac{4\pi\tilde G_0 \kappa^{-1} \alpha^2(\phi_0)(U + 
\tau - I^2)}{v \gamma}$ which is associated with scalar-tensor torsion.
We showed that the propagation of photons is unaffected by a screwed 
superconducting cosmic string and it is only affected by the angular 
deficit. This result shows us that the effect of torsion on massive 
particles is qualitatively different from its effect on radiation; this 
aspect becomes  especially relevant when calculating CMBR-anisotropy and 
the power spectrum as wiggly cosmic strings. One expects that this 
feature could help to partially by-pass the current difficulties in 
reconciling the COBE normalized matter power spectrum with the observational 
data in the cosmic string model.

\eject

{\bf Acknowledgments:} We would like to express our deep gratitude to
Prof.J.A. Helay\"el-Neto and Dr. J. H. Mosquera-Cuesta for helpful 
discussions on the subject of this paper.  We would like to 
thank (CNPq-Brazil) for financial support. We also thank Centro Brasileiro 
de Pesquisas F\'{\i}sicas (CBPF).

\end{document}